\documentclass[pre,twocolumn]{revtex4-1}
\usepackage{amsmath} 
\usepackage{graphicx}
\usepackage{color}

\begin{document}

\title{Single active particle in a harmonic potential: 
non-existence of the Jarzynski relation}
\author{Grzegorz Szamel}
\affiliation{Department of Chemistry, 
Colorado State University, Fort Collins, CO 80523}

\date{\today}

\begin{abstract}
The interest in active matter stimulates the need to generalize thermodynamic
description and relations to active matter systems, which are intrinsically out
of equilibrium. One important example is the Jarzynski relation, which 
links the exponential average of work done in an arbitrary
process connecting two equilibrium states with the difference of the free energies
of these states. Using a simple model system, a single thermal active Ornstein-Uhlenbeck
particle in a harmonic potential, we show that if the standard stochastic
thermodynamics definition of work is used, the Jarzynski relation is not generally
valid for processes between stationary states of active matter systems.
\end{abstract} 

\maketitle
\section{Introduction}
Active matter systems 
\cite{Ramaswamyrev1,Catesrev,Marchettirev1,Bechingerrev,Ramaswamyrev2,Marchettirev2}
consist of particles that consume energy from their environment to propel themselves. 
These systems are intrinsically out of equilibrium and therefore, as a matter of 
principle, standard relations derived for equilibrium systems do not apply to them.
Some of these relations can be generalized by introducing \textit{effective}
thermodynamic parameters, but, at least for now, there is no general framework
for doing this and thus the validity of such procedures has to be checked on a 
case by case basis. 

The Jarzynski relation \cite{Jarzynski1997} applies to processes connecting equilibrium
states of a system connected to a heat reservoir at temperature $T$ 
(here and in the following we set Boltzmann constant $k_B=1$). 
The Jarzynski relation states that the exponential average of the work $w$ done on 
the system while driving it between two equilibrium states is related to the 
difference of the free energies of these states,
\begin{equation}\label{Jarzynski1}
\left<\exp\left(- w/T \right)\right> = \exp\left(-\Delta F/T\right).
\end{equation} 
The beauty of relation \eqref{Jarzynski1} is that it is valid for an arbitrary
process connecting two given equilibrium states. Its usefulness is in that 
it allows to extract the free energy difference, \textit{i.e.} equilibrium 
information, from an ensemble of non-equilibrium trajectories of the system
\cite{Liphardt2002}. 

If one tries  to generalize the Jarzynski relation to processes
involving active matter systems, in principle one needs to generalize the notion
of the free energy. One can avoid this task by noticing that in the limit of 
infinitely slow processes one expects the work to become a non-fluctuating 
quantity. This allows one to replace $\Delta F$ by work done in an infinitely
slow, \textit{i.e.} quasistatic, process, $w^\text{qs}$. The generalized 
Jarzynski relation would
then connect the exponential average of the work done while driving
an active matter system between two stationary states to the work done in a quasistatic
process,
\begin{equation}\label{Jarzynski2}
\left<\exp\left(- w/T \right)\right> = \exp\left(-w^\text{qs}/T\right).
\end{equation} 
Here we investigate the existence of such a relation for a 
small active matter system using the standard stochastic thermodynamics definition
of work.

The problem with using the generalized Jarzynski relation to describe active matter 
systems is whether and how to generalize the 
temperature that prominently features in relation \eqref{Jarzynski2}. 
Different effective temperatures have been introduced for active matter systems
\cite{Loi2008,WangWolynes,Berthier2013,Szamel2014,Levis2015}
and it is \textit{a priori} not clear which one should be used if Eq. \eqref{Jarzynski2}
were to be extended to active matter systems. 

We study probably the simplest active matter system that can be externally
manipulated, a single \textit{thermal} active Ornstein-Uhlenbeck particle (AOUP) 
\cite{Szamel2014,Maggi2015,AOUPreview2020} 
in a harmonic potential. We consider two different classes of processes. In the 
first class we change the position of the minimum of the potential. In this case
no work is done in the quasistatic process and the right-hand-side of the 
Jarzynski equality becomes equal to 1. We show that in the limit of infinitely
fast and slow but finite speed 
processes two different effective temperatures have to be used
to keep the exponential average of the work equal to 1.  

In the second class of processes we change the force constant of the harmonic
potential. In this case, the quasistatic work is non-zero. We show that 
for both infinitely fast and slow but finite speed processes there is no effective
temperature that makes the generalized Jarzynski relation valid.

Our results demonstrate that at least with the standard definition of work
the Jarzynski relation is not generally valid for processes connecting stationary
states of active matter systems.

\section{A thermal AOUP in a harmonic potential}
We consider a single active particle moving in a harmonic potential. The particle
is endowed with a self-propulsion force that evolves according to the 
Ornstein-Uhlenbeck stochastic process. It also experiences the standard thermal
noise. The equations of motion read 
\begin{align}\label{eomx} 
\gamma\dot{x} & = - k \left(x-x_0\right) + f + \zeta &
\left< \zeta(t) \zeta(t')\right> & = 2 \gamma T \delta(t-t'), 
\\ \label{eomf}
\tau_p \dot{f} & = - f + \eta &
\left< \eta(t) \eta(t')\right>  & = 2 \gamma T_a \delta(t-t'). 
\end{align} 
In Eq. \eqref{eomx} $\gamma$ is the friction coefficient, $k$ is the force
constant, $x_0$ is the location of the potential minimum, 
$f$ is the self-propulsion and $\zeta$ is the thermal noise characterized
by temperature $T$.  
In Eq. \eqref{eomf} $\tau_p$ is the persistence time of the self-propulsion 
and $\eta$ is the noise of the reservoir coupled to the self-propulsion, 
characterized by active temperature $T_a$.  
Equivalently, the particle can be described by the Fokker-Planck equation
for a joint probability distribution of the position and the self-propulsion,
$P(x,f;t)$,
\begin{equation}\label{FP1}
\partial_t P(x,f;t) = \Omega P(x,f;t)
\end{equation}
where evolution operator $\Omega$ reads,
\begin{align}\label{Omega1}
\Omega = & - \gamma^{-1}\partial_x \left[-k \left(x-x_0\right) + f  
- T \partial_x \right]
\nonumber \\ & - \partial_f \left[-f/\tau_p 
-\left(\gamma T_a/\tau_p^2\right)\partial_f \right].
\end{align}
We note that the so-called drift coefficients \cite{VanKampen} in evolution
operator \eqref{Omega1}
are linear in $x$ and $f$ and therefore the stationary solution of Eq. \eqref{FP1} 
has a Gaussian form. In particular, the position distribution reads
\begin{equation}\label{posss}
p^\text{ss}(x) \propto \exp\left[-\frac{\left(1/2\right) k \left(x-x_0\right)^2}
{T + T_a/\left(k\tau_p/\gamma+1\right)}\right].
\end{equation}
We follow standard stochastic thermodynamics \cite{Sekimoto1998,Seifert2012} and
define the work done while changing parameter $\alpha$ of the potential 
$U(x) = (1/2) k (x-x_0)^2$ as
\begin{equation}\label{work1} 
w = \int_0^\tau dt \, \dot{\alpha} \, \partial_\alpha U(x).
\end{equation}
We consider two classes of processes, with $\alpha=x_0$ and $\alpha=k$. 

\section{Work done by shifting the potential minimum} 
In the case of moving the potential minimum, $x_0\to x_0+\Delta x_0$, 
the work done in an infinitely slow (quasi-static) process vanishes,
$w^\text{qs}=0$, and,
as stated earlier, the right-hand-side of generalized Jarzynski relation,
Eq. \eqref{Jarzynski2}, is equal to 1. 

The work done in an instantaneous process is equal to 
$w^\text{ins} = (1/2)k\left(- 2 \Delta x_0 (x-x_0)+\Delta x_0^2\right)$
and its distribution reads
\begin{equation}\label{workins1}
p^\text{ins}(w) = \left< \delta\left(w
- (1/2)k\left(- 2 \Delta x_0 (x-x_0)+\Delta x_0^2\right)\right)
\right>^\text{ss}
\end{equation}
where here and in the following $\left<\ldots\right>^\text{ss}$ 
denotes averaging over the
stationary distribution. Explicit calculation shows that in this case the 
generalized Jarzynski relation is satisfied with 
$T_{eff 1}=T + T_a/\left(k\tau_p/\gamma+1\right)$,
\begin{equation}\label{Jarzynskiins1}
\left<\exp\left(-w/T_{eff 1}\right)\right>^\text{ins} = 1,
\text{\hskip 1em} T_{eff 1}=T + T_a/\left(k\tau_p/\gamma+1\right),
\end{equation}
where $\left<\ldots\right>^\text{ins}$ denotes averaging over distribution
of work done in an instantaneous process. We note that effective temperature 
\eqref{Jarzynskiins1} is the temperature that is obtained if stationary 
state distribution of particle positions, \eqref{posss}, 
is interpreted as the Gibbs measure,
$p^\text{ss}(x) \propto \exp\left(-U(x)/T_{eff 1}\right)$. 
Furthermore, Eq. \eqref{Jarzynskiins1} is consistent with
the result of Paneru \textit{et al.} \cite{Paneru} who considered work
extracted from an active information engine. 

For finite-speed processes it is convenient to follow 
Mazonka and Jarzynski \cite{Mazonka} and write a Fokker-Planck
equation for a joint probability distribution for the position, self-propulsion
and work, $p(x,f,w;t)$, 
\begin{equation}\label{FP2}
\partial_t p(x,f,w;t) = \left[\Omega + \dot{x}_0 k (x-x_0)\partial_w \right] p(x,f,w;t).
\end{equation}
Assuming that at the start of driving the particle is in the stationary state
we get the following initial condition for Eq. \eqref{FP2}, 
$p(x,f,w;t=0)=p^\text{ss}(x,f) \delta(w)$.
Once again we note that drift coefficients in Eq. \eqref{FP2} are linear. This fact
and a Gaussian (albeit singular) initial condition $p(x,f,w;t=0)$ imply that
distribution $p(x,f,w;t)$ is a Gaussian distribution 
with time-dependent coefficients. It follows
that work distribution $p(w;t)=\int dx df\, p(x,f,w;t)$ is also a Gaussian 
and therefore it is fully characterized by the first two cumulants of the work. 
To calculate these cumulants we use Eq. \eqref{FP2} and for slow but finite speed 
driving we get the following result (see Appendix A for details of the calculation)
\begin{align}\label{workslow1}
\left<w\right>^\text{sl} & = \dot{x}_0 \gamma \Delta x_0,
\\  \label{workvarslow1}
\sigma^2_w= \left<w^2\right>^\text{sl}-\left(\left<w\right>^\text{sl}\right)^2 & 
= 2 \dot{x}_0 \gamma \Delta x_0 \left(T+T_a\right) ,
\end{align}
where $\left< \ldots \right>^\text{sl}$ denotes averaging over the work 
distribution for slow but finite speed driving. In the limit of infinitely
slow driving, \textit{i.e.} in the quasi-static limit $\dot{x}_0\to 0$, 
the variance of the work vanishes,
\textit{i.e.} the work does not fluctuate, and the average work vanishes as well. 

For a Gaussian distribution of work generalized Jarzynski relation 
\eqref{Jarzynski2} is 
satisfied with $T$ replaced by effective temperature $T_{eff}$ if \cite{Mazonka} 
\begin{equation}\label{Gaussian}
\left<w\right> = w^\text{qs} + \sigma^2_w/\left(2 T_{eff}\right).
\end{equation}
Thus, results (\ref{workslow1}-\ref{workvarslow1}) imply that 
for slow but finite speed driving the generalized Jarzynski relation is satisfied with 
$T_{eff 2}=T + T_a$,
\begin{equation}\label{Jarzynskislow1}
\left<\exp\left(-w/T_{eff 2}\right)\right>^\text{sl} = 1,
\text{\hskip 2em} T_{eff 2} = T + T_a
\end{equation}
(recall that $w^\text{qs}=0$ for moving the potential minimum). 
We note that the effective temperature that enters into the generalized 
Jarzynski relation for slow but finite speed driving is the same as the effective temperature
obtained from the fluctuation-dissipation ratio in the limit of small frequencies
(see Appendix \ref{ap:FDTTeff} for details).

We emphasize that the fact that two different effective temperatures are required 
to make the generalized Jarzynski relation valid for this very simple class of 
processes implies that even for a thermal AOUP in a harmonic potential effective 
temperature is a non-unique notion and there is no ``the effective temperature''.
 
\section{Work done by changing the force constant}
Next, we consider work done by increasing the force constant of the 
potential, $k\to k+\Delta k$. To simplify the notation in this section
we set the potential minimum at $x_0=0$. 
For the increase of the force constant, the work done in an infinitely
slow (quasistatic) process is non-zero. The quasistatic work can be calculated by 
rewriting Eq. \eqref{work1} with $\alpha=k$ as an integration over k 
of $\partial_k U(x) = \frac{x^2}{2}$ averaged over stationary state distribution, Eq. \eqref{posss}, 
\begin{equation}\label{workqs2}
w^\text{qs} = \int_k^{k+\Delta k} dk_1\, \left<\frac{x^2}{2}\right>^\text{ss}.
\end{equation}
The result reads
\begin{equation}\label{workqs3}
w^\text{qs} = 
\frac{T+T_a}{2}\ln\frac{k+\Delta k}{k} 
- \frac{T_a}{2}
\ln\frac{(k+\Delta k)\tau_p/\gamma+1}{k\tau_p/\gamma+1}.
\end{equation}

The work done in an instantaneous process is equal to 
$w^\text{ins} = (1/2)\Delta kx^2$
and its distribution reads
\begin{align}\label{workins2}
p^\text{ins}(w) & = \left< \delta\left(w- (1/2)\Delta kx^2\right)\right>^\text{ss}
\\ \nonumber & =
\sqrt{\frac{k}{\pi w \Delta k T_{eff 1}}}
e^{-\frac{kw}{\Delta k T_{eff 1}}},
\end{align}
where $T_{eff 1}$ is defined in Eq. \eqref{Jarzynskiins1}. 
Explicit calculation shows that for instantaneous changes of the force constant 
$T_{eff 1}$ \textit{cannot} be used in the generalized Jarzynski relation,
\begin{equation}\label{Jarzynskiins2}
\left<\exp\left(-w/T_{eff 1}\right)\right>^\text{ins} = 
\sqrt{\frac{k+\Delta k}{k}}
\neq
\exp\left(-w^\text{qs}/T_{eff 1}\right).
\end{equation}
In fact, there is no effective temperature that is independent of the 
change of the force constant and that leads to the generalized Jarzynski relation for
work distribution \eqref{workins2}.

To investigate the existence of the generalized Jarzynski relation for
a slow but finite rate increase of the force constant we derive an approximate
distribution of work done in this process. To this end we follow 
Speck \cite{Speck2011} who derived the analogous distribution for work
done on a Brownian particle in a harmonic potential. 
The calculation is somewhat tedious but straightforward; it is presented in
Appendix \ref{ap:kslow}. The approximate distribution of work for a slow but finite 
speed change of the force constant is a Gaussian with cumulants that are given by 
the following, rather complicated expressions,
\begin{align}\label{workslow2}
\left<w\right>^\text{sl} & = w^\text{qs} + \dot{k} \int_k^{k+\Delta k} dk_1\, 
\left[\frac{\gamma T}{4k_1^3}
\right. \nonumber \\ & \left.
+
\frac{\gamma T_a
\left(4\left(k_1\tau_p/\gamma\right)^2+3 k_1\tau_p/\gamma+1\right)}
{4k_1^3 \left(k_1\tau_p/\gamma+1\right)^3}\right]
\\  \label{workvarslow2}
\sigma^2_w 
& = 2 \dot{k} \int_k^{k+\Delta k} dk_1\, \left[ \frac{\gamma T^2}{4k_1^3} 
+ \frac{\gamma T T_a \left(2k_1\tau_p/\gamma+1\right)}
{2k_1^3\left(k_1\tau_p/\gamma+1\right)^2}
\right. \nonumber \\ & \left.
+ \frac{\gamma T_a^2
\left(\left(k_1\tau_p/\gamma\right)^2+3k_1\tau_p/\gamma+1\right)}
{4k_1^3\left(k_1\tau_p/\gamma+1\right)^3}\right]
\end{align}
Cumulants (\ref{workslow2}-\ref{workvarslow2}) 
do not satisfy relation \eqref{Gaussian} and therefore, again, 
there is no effective temperature that leads to the generalized Jarzynski relation for 
a slow but finite speed process in which the force constant is increased.

\section{Discussion}
Our results imply that if one uses the standard definition of work, 
Jarzynski relation \eqref{Jarzynski1} generally cannot 
be extended to active matter systems. For some classes of processes,
depending on the speed of driving different
effective temperatures have to be used to make generalized Jarzynski relation 
\eqref{Jarzynski2} valid. For other classes of 
processes there is no effective temperature that would lead to the 
generalized Jarzynski relation.

We emphasize that this result follows if one uses the definition of work
utilized in standard stochastic thermodynamics, Eq. \eqref{work1}. 
It is possible that other definitions of work, 
\textit{e.g.} excess work defined by Hatano and Sasa \cite{HatanoSasa}, could 
lead to a generalized Jarzynski relation. This may seem plausible since although 
in this work we showed that the Jarzynski relation generally cannot be extended to
active matter systems, elsewhere \cite{Szamel2019} we showed that
fluctuation theorems for different kinds of entropy are in general satisfied 
for active matter systems \cite{Chaudhuri}. On the other hand, we recall that fluctuation
theorems for entropy do not involve the temperature and thus the issue
that makes the generalized Jarzynski relation invalid does not occur.  
This subject is left for a future investigation. 

Our finding is consistent with the fact that in out-of-equilibrium systems 
if one uses relations that give thermodynamic temperature for equilibrium systems, 
one generally gets different effective temperatures. Simply speaking, there is no 
``the effective temperature''. Only in certain cases, \textit{e.g.}
for glassy systems under shear, \textit{a priori} different effective temperatures
turn out to have the same value \cite{Berthier2002a,Berthier2002b}. This result
follows from a well understood theoretical argument involving the separation 
of time scales of different relaxation processes. 

\section*{Acknowledgments}

I thank Elijah Flenner 
for comments on the manuscript. I gratefully acknowledge the support
of NSF Grant No.~CHE 2154241.

\appendix

\section{Work done while shifting the potential minimum}\label{ap:moments}

Our calculation of the first two moments of the distribution of the 
work done while shifting the potential minimum follows a similar
calculation presented in Ref. \cite{Mazonka}, which was concerned with 
a (passive) Brownian particle. 

We consider a thermal AOUP in a harmonic potential. To simplify the notation
in this appendix we set the initial potential minimum at $x_0(t=0)=0$. 

We assume that at time 
$t=0$ the potential minimum starts moving with constant velocity $\dot{x}_0$.
Using Eq. \eqref{work1} we obtain the following equation of motion for the work done 
while shifting the potential minimum,
\begin{equation}\label{eomwA}
\partial_t w = - \dot{x}_0 k \left(x-\dot{x}_0 t\right).
\end{equation}
Following Ref. \cite{Mazonka} we switch to the reference frame of the moving
potential minimum and introduce a new variable, $y=x-\dot{x}_0 t$. Next, we
re-write Fokker-Planck equation \eqref{FP2} of the main text as an equation 
describing the joint probability distribution
for the particle's position (in the reference frame of the moving
potential minimum), self-propulsion and work accumulated between the initial 
time and time $t$, 
\begin{align}\label{FPA1} 
& \partial_t p(y,f,w;t) = \left\{ 
- \gamma^{-1}\partial_y \left[-k y -\gamma \dot{x}_0 + f  
- T \partial_y \right] 
\right. \nonumber \\ & \left.
- \partial_f \left[-f/\tau_p 
-\left(\gamma T_a/\tau_p^2\right)\partial_f \right] + \dot{x}_0 k y \partial_w  
\right\}  p(y,f,w;t).
\end{align}
The initial condition for Eq. \eqref{FPA1} reads 
$p(y,f,w;t=0)=p^\text{ss}(y,f) \delta(w)$.

Starting from Eq. \eqref{FPA1} we can derive the following equations
for the averages of $w$ and $y$, and $f$,
\begin{align}\label{eomaw}
\partial_t \left<w\right> = & - \dot{x}_0 k \left<y\right>,
\\ \label{eomay}
\partial_t \left<y\right> = & - \dot{x}_0 - \left(k/\gamma\right) \left<y\right>
+ \gamma^{-1}\left<f\right>,
\\ \label{eomaf}
\partial_t \left<f\right> = & - \tau_p^{-1} k \left<f\right>.
\end{align}
Initial conditions for these equation are $\left<w\right>(t=0)=0$,
$\left<y\right>(t=0)=0$ and $\left<f\right>(t=0)=0$. 

Solving Eqs. (\ref{eomaw}-\ref{eomaf}) we get 
$\left<w\right>(t) = \dot{x}_0^2 \gamma t + \left(\dot{x}_0^2/k\right)
\left(\exp\left(-kt/\gamma\right)-1\right)$. Recalling that the change of the
potential minimum over time $\tau$ is $\Delta x_0\equiv \dot{x}_0 \tau$,
we obtain
\begin{equation}\label{avew}
\left<w\right> = \dot{x}_0 \gamma \Delta x_0  + 
\left(\dot{x}_0^2/k\right)\left(\exp\left(-k\Delta x/\gamma\dot{x}_0\right)-1\right).
\end{equation}
In the limit of slow but finite driving Eq. \eqref{avew} leads to Eq. \eqref{workslow1}.
Incidentally, in the limit of infinitely fast process Eq. \eqref{avew} reproduces
$\left<w\right>^\text{ins} = k\Delta x_0^2/2$.

To evaluate the variance of the work we need to derive equations of motions
for the second cumulants. We follow the notation of Ref. \cite{Mazonka}, 
\begin{align}\label{cumdef}
\sigma^2_w & =\left<w^2\right>-\left<w\right>^2 ,
\\ 
c_{yw} & = \left<yw\right>-\left<y\right>\left<w\right>,
\end{align}
\textit{etc.} The derivation of equations of motion is straightforward but somewhat 
lengthy. The result is 
\begin{align}\label{eomsw}
\partial_t \sigma^2_w & = - 2\dot{x}_0 k c_{yw},
\\ \label{eomcyw}
\partial_t c_{yw} & = - \left(k/\gamma\right)c_{yw} + \gamma^{-1}c_{fw} 
-\dot{x}_0 k \sigma^2_y,
\\ \label{eomsy}
\partial_t \sigma^2_y & = - \left(2k/\gamma\right) \sigma^2_y 
+ \left(2/\gamma\right) c_{yf},
\\ \label{eomcyf}
\partial_t c_{yf} & = - \left(k/\gamma + 1/\gamma\right)c_{yf} + \gamma^{-1}\sigma^2_f ,
\\ \label{eomsf}
\partial_t \sigma^2_f & = - \left(2/\tau_p\right) \sigma^2_f 
+ \left(2\gamma T_a/\tau_p^2\right),
\\ \label{eomcfw}
\partial_t c_{fw} & = - \tau_p^{-1} c_{fw} -\dot{x}_0 k c_{yf}.
\end{align}
We note that Eqs. (\ref{eomsy}-\ref{eomsf}) do no couple to the other equations;
since initial conditions for Eqs. (\ref{eomsy}-\ref{eomsf}) are stationary state
averages, $\sigma^2_y$, $c_{yf}$ and $\sigma^2_f$ will not change,
\begin{align}\label{sy}
\sigma^2_y & = T/k+T_a/\left(k\left(k\tau_p/\gamma+1\right)\right), 
\\ \label{cyf}
c_{yf} & = T_a/\left(k\tau_p/\gamma+1\right),
\\ \label{sf}
\sigma^2_f &= \gamma T_a/\tau_p.
\end{align}
The remaining equations can be integrated, starting from Eq. \eqref{eomcfw},
then moving to \eqref{eomcyw} and finally \eqref{eomsw}. Then we again
recall that the change of the
potential minimum over time $\tau$ is $\Delta x_0\equiv \dot{x}_0 \tau$ and we get
\begin{align}\label{sw}
\sigma^2_w & = 2 \dot{x}_0 k \Delta x_0\left(\gamma \sigma^2_y+\tau_p c_{yf}\right)
\nonumber \\ & + 2\dot{x}_0^2 \left(\gamma^2 \sigma^2_y
+ \frac{\gamma\tau_p}{k\tau_p/\gamma-1} c_{yf}\right)
\left(\exp\left(-k\Delta x/\gamma\dot{x}_0\right)-1\right)
\nonumber \\ & 
+  2\dot{x}_0^2 
\frac{k^2\tau_p^3}{\gamma\left(k\tau_p/\gamma-1\right)}
\left(\exp\left(-\Delta x/\tau_p\dot{x}_0\right)-1\right)
\end{align}
In the limit of slow but finite driving Eq. \eqref{sw} gives Eq. \eqref{workvarslow1}.
Once again, it can be shown that in the limit of infinitely fast process Eq. \eqref{sw}
reproduces the variance of distribution $p^\text{ins}$, Eq. \eqref{workins1}.

\section{Fluctutation-dissipation ratio-based effective temperature}\label{ap:FDTTeff}

The calculation outlined in this Appendix generalizes that presented in Sec. IV 
of Ref. \cite{Szamel2014} for a single \textit{athermal} AOUP in a harmonic potential.
To simplify the notation, in this appendix we set the potential minimum
at $x_0=0$.

Following Ref. \cite{Cugliandolo} we define a frequency-dependent
fluctuation-dissipation ratio-based effective temperature 
\begin{equation}\label{Tefffdrdef}
T_{eff}^{FDR}(\omega) = \frac{\omega \mathrm{Re } C(\omega)}{\chi''(\omega)},
\end{equation}
where $\mathrm{Re } C(\omega)$ is the real part of the 
one-sided Fourier transform of the particle's position
auto-correlation function, 
$\mathrm{Re } C(\omega) = \mathrm{Re } 
\int_0^\infty e^{i\omega t} \left<x(t)x(0)\right>$,
and $\chi''(\omega)$ is the imaginary part of the one-sided Fourier transform of the 
response function, $\chi''(\omega) = \mathrm{Im }\int_0^\infty e^{i\omega t} R(t)$,
where $R(t)$ describes the change of the particle's position due to an external force. 

To calculate the position auto-correlation function we start from
equations of motion (\ref{eomx}-\ref{eomf}) 
and derive the following set of coupled equations 
for $\left<x(t)x(0)\right>$ and $\left<f(t)x(0)\right>$,
\begin{eqnarray}\label{eomcf1}
\gamma \partial_t \left<x(t)x(0)\right> &=& - k \left<x(t)x(0)\right> 
+ \left<f(t)x(0)\right> 
\\ \label{eomcf2}
\tau_p \partial_t \left<f(t)x(0)\right> &=& - \left<f(t)x(0)\right>.
\end{eqnarray} 
Since thermal noise in uncorrelated with the initial position of the particle,
the above equations have the same form as those in Sec. IVB of Ref. \cite{Szamel2014}.

The presence of the thermal noise influences the initial conditions for 
Eqs. (\ref{eomcf1}-\ref{eomcf2}),
\begin{align}\label{eomcf1init}
\left<x(0)x(0)\right> & \equiv  \left<x^2\right>
= \frac{T}{k} 
+ \frac{T_a}{k\left(k\tau_p/\gamma+1\right)},
\\ \label{eomcf2init}
\left<f(0)x(0)\right> &\equiv  \left<fx\right>=k\left<x^2\right> - T = 
\frac{T_a}{k\tau_p/\gamma+1}.
\end{align} 

Equations of motion (\ref{eomcf1}-\ref{eomcf2}) with initial conditions
(\ref{eomcf1init}-\ref{eomcf2init}) lead to the following expression
for the position auto-correlation function,
\begin{align}\label{acf}
\left<x(t)x(0)\right> = & \frac{T_a/k}{\left(k\tau_p/\gamma\right)^2-1}
\left(\left(k\tau_p/\gamma\right)e^{-t/\tau_p} - e^{-k t/\gamma}\right)
\nonumber \\ & 
+ \left(T/k\right)e^{-k t/\gamma}.
\end{align}

To evaluate the response function we add to Eq. \eqref{eomx} a weak, time-dependent 
external force $F^\text{ext}(t)$ and then derive coupled equations of motion
for the resulting change of the average position of the particle and of the
self-propulsion
\begin{align}\label{eomlrx}
\gamma \partial_t \delta \left<x(t)\right> &= \delta \left<f(t)\right> 
- k\delta \left<x(t)\right> + F^\text{ext}(t)
\\ \label{eomlrf}
\tau_p \partial_t \delta\left<f(t)\right> &= -\delta\left<f(t)\right>.
\end{align}
The initial conditions for these equations are 
$\delta \left<x(t=0)\right> = 0 = \delta \left<f(t=0)\right>$.

Solving Eqs. (\ref{eomlrx}-\ref{eomlrf}) we get $\delta \left<f(t)\right>\equiv 0$
and 
\begin{equation}\label{xlr}
\delta \left<x(t)\right> = 
\frac{1}{\gamma}\int_0^t dt' e^{-k(t-t')/\gamma} F^\text{ext}(t').
\end{equation}
The response function thus is given by 
\begin{equation}\label{lrf}
R(t)=\left(1/\gamma\right) e^{-kt/\gamma}.
\end{equation}
Response function \eqref{lrf} is the same as that derived in Ref. \cite{Szamel2014}
for an athermal AOUP.

Using Eqs. \eqref{acf} and \eqref{lrf} we get from Eq. \eqref{Tefffdrdef}
\begin{equation}\label{Tefffdr}
T_{eff}^{FDR}(\omega) = \frac{T_a}{1+\tau_p^2\omega^2} + T
\end{equation}
In the small frequency limit $T_{eff}^{FDR}(\omega)$ becomes $T_a+T$ and thus it
coincides with the effective temperature that makes Jarzynski relation valid for 
the work distribution in a slow but finite shift of the potential minimum.
We note that $T_a+T$ is also the effective temperature that is obtained 
from the long-time diffusion coefficient of a free thermal AOUP.

\section{Work distribution for slow but finite increase of the force constant}\label{ap:kslow}

Our calculation of the approximate distribution of the work done while increasing
the force constant follows a similar calculation presented in Ref. \cite{Speck2011},
which was concerned with a (passive) Brownian particle. 

We consider a thermal AOUP in a harmonic potential. To simplify the notation
in this appendix we set the potential minimum at $x_0=0$. 

We assume that at time 
$t=0$ the force constant starts increasing with constant velocity $\dot{k}$.
Using Eq. \eqref{work1} we obtain the following equation of motion for the work done 
while increasing the force constant,
\begin{equation}\label{eomwC}
\partial_t w = \frac{\dot{k}}{2} x^2.
\end{equation}
Next, we write a evolution equation that is similar to Eq. \eqref{FP2}, which
describes the time dependence of the joint probability distribution
for the particle's position, self-propulsion and work accumulated between the initial 
time and time $t$, 
\begin{align}\label{FPC1} 
& \partial_t p(x,f,w;t) = \left\{ 
- \gamma^{-1}\partial_x \left[-k x + f  
- T \partial_x \right] 
\right. \\ & \nonumber \left.
- \partial_f \left[-f/\tau_p 
-\left(\gamma T_a/\tau_p^2\right)\partial_f \right] 
- \left(\dot{k}x^2/2\right)\partial_w  
\right\}  p(x,f,w;t).
\end{align}
The initial condition for Eq. \eqref{FPC1} is 
$p(x,f,w;t=0)=p^\text{ss}(x,f) \delta(w)$.

We note that one of the drift coefficients in Eq. \eqref{FPC1} is quadratic
and thus the time-dependent distribution $p(x,f,w;t)$ does not have Gaussian form.
In fact, since for a process in which the force increases the time derivative of the 
work, Eq. \eqref{eomwC}, is always positive, $p(x,f,w;t)=0$ for $w<0$ and thus
distribution $p(x,f,w;t)$ cannot be a Gaussian.

However, if we introduce characteristic function \cite{VanKampen},
\begin{equation}\label{rho}
\rho(x,f,\lambda;t) = \int dw e^{i\lambda w} p(x,f,w;t),
\end{equation}
we note that the equation of motion for $\rho(x,f,\lambda;t)$
\begin{align}\label{FPC2} 
& \partial_t \rho(x,f,\lambda;t) = \left\{ 
- \gamma^{-1}\partial_x \left[-k x + f  
- T \partial_x \right] 
\right. \\ & \nonumber \left.
- \partial_f \left[-f/\tau_p 
-\left(\gamma T_a/\tau_p^2\right)\partial_f \right] + i \lambda\dot{k}x^2/2  
\right\}  \rho(x,f,\lambda;t)
\end{align}
allows for a solution that has a Gaussian form \cite{Speck2011}. 
This fact allows us to derive a closed set of equations describing the time
dependence of the characteristic function of the work distribution,
\begin{equation}\label{psi}
\psi(\lambda;t) = \int dx df \rho(x,f,\lambda;t).
\end{equation}
The first equation expresses the time derivative of $\psi(\lambda;t)$ in terms of 
the generalized second moment,
\begin{equation}\label{psieom1}
\partial_t \psi(\lambda;t) = 
i\lambda \dot{k} \phi_{x^2}(\lambda;t)/2
\end{equation}
where the generalized second moment $\phi_{x^2}(\lambda;t)$ reads
\begin{eqnarray}\label{phix2}
\phi_{x^2}(\lambda;t) =  \int dx df\, x^2 \rho(x,f,\lambda;t).
\end{eqnarray}
The generalized second moment satisfies the following equation
\begin{align}\label{phix2eom1}
\partial_t \phi_{x^2}(\lambda;t) &=
\frac{2}{\gamma}\left(\phi_{xf}(\lambda;t) - k \phi_{x^2}(\lambda;t)\right)
+\frac{2T}{\gamma} \psi(\lambda;t)
\nonumber \\ & 
+\frac{i\lambda\dot{k}}{2} \int dx df\, x^4 \rho(x,f,\lambda;t),
\end{align}
where $\phi_{xf}(\lambda;t)$ denotes the mixed generalized second moment,
\begin{eqnarray}\label{phixf}
\phi_{xf}(\lambda;t) =  \int dx df\, xf \rho(x,f,\lambda;t).
\end{eqnarray}
To close the equations of motion we will also need $\phi_{f^2}(\lambda;t)$,
\begin{eqnarray}\label{phif2}
\phi_{f^2}(\lambda;t) =  \int dx df\, f^2 \rho(x,f,\lambda;t)
\end{eqnarray}
Equations of motion for $\phi_{xf}(\lambda;t)$ and $\phi_{f^2}(\lambda;t)$ read
\begin{align}\label{phixfeom1}
\partial_t \phi_{xf}(\lambda;t) & = 
\frac{1}{\gamma}\phi_{f^2}(\lambda;t)
- \left(\frac{k}{\gamma}+\frac{1}{\tau_p}\right) \phi_{xf}(\lambda;t)
\nonumber \\ & 
+\frac{i\lambda\dot{k}}{2} \int dx df\, x^3 f \rho(x,f,\lambda;t),
\end{align}
\begin{align}\label{phif2eom1}
\partial_t \phi_{f^2}(\lambda;t) & = 
- \frac{2}{\tau_p}\phi_{f^2}(\lambda;t)
+ \frac{2 \gamma T_a}{\tau_p^2} \psi(\lambda;t)
\nonumber \\ & 
+\frac{i\lambda\dot{k}}{2} \int dx df x^2 f^2 \rho(x,f,\lambda;t).
\end{align}
As noted above, equation of motion for $\rho(x,f,\lambda;t)$ allows for 
a solution that has a Gaussian form. However, due to the last term at the
right-hand-side of Eq. \eqref{FPC2}, this Gaussian distribution is not
normalized. The un-normalized Gaussian form of $\rho(x,f,\lambda;t)$ 
allows us to express higher-order moments in terms of generalized second moments,
\begin{align}\label{phix4}
\int dx df\, x^4 \rho(x,f,\lambda;t) = \frac{3 \phi_{x^2}^2(\lambda;t)}{\psi(\lambda;t)}
\end{align}
\begin{align}\label{phix3f}
\int dx df\, x^3 f \rho(x,f,\lambda;t) = 
\frac{3\phi_{x^2}(\lambda;t)\phi_{xf}(\lambda;t)}{\psi(\lambda;t)}
\end{align}
\begin{align}\label{phix2f2}
\int dx df\, x^2 f^2 \rho(x,f,\lambda;t) =
\frac{\phi_{x^2}(\lambda;t)\phi_{f^2}(\lambda;t)}{\psi(\lambda;t)}
+\frac{2 \phi_{xf}^2(\lambda;t)}{\psi(\lambda;t)}
\end{align}
Using closures (\ref{phix4}-\ref{phix2f2}) in the equations of motion for the
generalized moments we get the following closed set of equations,
\begin{align}\label{phix2eom2}
\partial_t \phi_{x^2}(\lambda;t) & = 
\frac{2}{\gamma}\left(\phi_{xf}(\lambda;t) - k \phi_{x^2}(\lambda;t)\right)
+\frac{2T}{\gamma} \psi(\lambda;t)
\nonumber \\ &
+\frac{3i\lambda\dot{k}}{2}\frac{\phi_{x^2}^2(\lambda;t)}{\psi(\lambda;t)},
\\ \label{phixfeom2}
\partial_t \phi_{xf}(\lambda;t) &= 
\frac{1}{\gamma}\phi_{f^2}(\lambda;t)
- \left(\frac{k}{\gamma}+\frac{1}{\tau_p}\right) \phi_{xf}(\lambda;t)
\nonumber \\ &
+\frac{3i\lambda\dot{k}}{2}\frac{\phi_{xf}(\lambda;t)\phi_{x^2}(\lambda;t)}
{\psi(\lambda;t)},
\\ \label{phif2eom2}
\partial_t \phi_{f^2}(\lambda;t) &=
- \frac{2}{\tau_p}\phi_{f^2}(\lambda;t)
+ \frac{2 \gamma T_a}{\tau_p^2} \psi(\lambda;t)
\nonumber \\ &
+\frac{i\lambda\dot{k}}{2}
\frac{\phi_{x^2}(\lambda;t)\phi_{f^2}(\lambda;t)}{\psi(\lambda;t)}
+i\lambda\dot{k}\frac{\phi_{xf}^2(\lambda;t)}{\psi(\lambda;t)}
\end{align}

Next, following Ref. \cite{Speck2011} we assume constant rate of change of
the force constant, $\dot{k}=\text{const.}$ and expand generalized second moments
in powers of $\dot{k}$,
\begin{align}\label{phix2e1}
\phi_{x^2}(\lambda;t) &= \phi_{x^2}^{(0)}(\lambda;t) + \dot{k}\phi_{x^2}^{(1)}(\lambda;t)
+ \ldots ,
\\ \label{phixfe1}
\phi_{xf}(\lambda;t) &= \phi_{xf}^{(0)}(\lambda;t) +  \dot{k}\phi_{xf}^{(1)}(\lambda;t)
+ \ldots ,
\\ \label{phif2e1}
\phi_{x^2}(\lambda;t) &= \phi_{f^2}^{(0)}(\lambda;t) + \dot{k}\phi_{f^2}^{(1)}(\lambda;t)
+ \ldots .
\end{align}
We substitute expansions (\ref{phix2e1}-\ref{phif2e1}) into the equations of motion.
We also assume that the time derivatives are of order $\dot{k}$. In this way we 
get the following set of equations for the zeroth order terms,
\begin{align}\label{phix2eom3}
0 &=
\frac{2}{\gamma}\left(\phi_{xf}^{(0)}(\lambda;t) - k \phi_{x^2}^{(0)}(\lambda;t)\right)
+\frac{2T}{\gamma} \psi(\lambda;t),
\\ \label{phixfeom3}
0 &= 
\frac{1}{\gamma}\phi_{f^2}^{(0)}(\lambda;t)
- \left(\frac{k}{\gamma}+\frac{1}{\tau_p}\right) \phi_{xf}^{(0)}(\lambda;t),
\\ \label{phif2eom3}
0 &= 
- \frac{2}{\tau_p}\phi_{f^2}^{(0)}(\lambda;t)
+ \frac{2 \gamma T_a}{\tau_p^2} \psi(\lambda;t).
\end{align}
Solving these equations we get the following result for 
$\phi_{x^2}^{(0)}(\lambda;t)$,
\begin{equation}\label{phix2e2}
\phi_{x^2}^{(0)}(\lambda;t) = \frac{T}{k}\psi(\lambda;t) 
+ \frac{T_a}{k\left(k\tau_p/\gamma + 1\right)}\psi(\lambda;t).
\end{equation}
We use result \eqref{phix2e2} in Eq. \eqref{psieom1} and get the following result
for $\psi(\lambda;t)$,
\begin{align}\label{psiex0}
\psi(\lambda;t) & = \exp\left\{\frac{i\lambda}{2}
\left[\left(T+T_a\right)\ln\frac{k(t)}{k(0)}
\right.\right.  \nonumber \\ & \left.\left.
- T_a\ln\frac{k(t)\tau_p/\gamma+1}{k(0)\tau_p/\gamma+1}\right]\right\}.
\end{align}
Equation \eqref{psiex0} means that the work distribution is a delta function
centered at the quasistatic work given by Eq. \eqref{workqs3}.  

Next, we consider terms of order $\dot{k}$ in the equations of motion. After
some manipulations we get 
\begin{align}\label{phixfeom4}
\phi_{x^2}^{(1)}(\lambda;t) &= 
\left[\frac{\gamma T}{2k^3}
\right. \nonumber \\ & \left.
+
\frac{\gamma T_a
\left(2\left(k\tau_p/\gamma\right)^2+3 k\tau_p/\gamma+1\right)}
{2k^3 \left(k\tau_p/\gamma+1\right)^3}\right]\psi(\lambda;t)
\nonumber \\ &
+ i\lambda \left[ \frac{\gamma T^2}{2k^3} 
+ \frac{\gamma T T_a \left(2k\tau_p/\gamma+1\right)}
{k^3\left(k\tau_p/\gamma+1\right)^2}
\right. \nonumber \\ & \left.
+ \frac{\gamma T_a^2
\left(\left(k\tau_p/\gamma\right)^2+3k\tau_p/\gamma+1\right)}
{2k^3\left(k\tau_p/\gamma+1\right)^3}\right]\psi(\lambda;t)
\end{align}
Using the right-hand-side of Eq. \eqref{phixfeom4} in Eq. \eqref{psieom1} we 
see that at the first order in $\dot{k}$ the work distribution is a Gaussian
with the first cumulants given by expressions (\ref{workslow2}-\ref{workvarslow2}).

In closing we note that the Gaussian form of the work distribution is only an 
approximation, since the true distribution vanishes for $w<0$ and therefore cannot 
have Gaussian form.

\end{document}